\documentclass[%
twocolumn,
showpacs,
 amsmath,amssymb,
 aps,prc,
superscriptaddress,
]{revtex4-1}

\usepackage{siunitx}
\usepackage{graphicx}
\usepackage{dcolumn}
\usepackage{bm}
\usepackage{isotope}

\newcolumntype{C}[1]{>{\centering\arraybackslash}m{#1}}
\begin{document}


\title{Measurement of the $^{187}$Re($\alpha$,n)$^{190}$Ir reaction cross section at sub-Coulomb energies using the Cologne Clover Counting Setup}

\author{P.~Scholz}
\email[]{pscholz@ikp.uni-koeln.de}
\affiliation{Institute for Nuclear Physics, University of Cologne,  50937 K\"oln, Germany}
\author{A.~Endres}
\altaffiliation[Present address: ]{Institut f\"ur angewandte Physik, Goethe-Universit\"at Frankfurt am Main, 60438 Frankfurt am Main, Germany}
\affiliation{Institute for Nuclear Physics, University of Cologne,  50937 K\"oln, Germany}
\author{A.~Hennig}
\affiliation{Institute for Nuclear Physics, University of Cologne,  50937 K\"oln, Germany}
\author{L.~Netterdon}
\affiliation{Institute for Nuclear Physics, University of Cologne,  50937 K\"oln, Germany}
\author{H.W.~Becker}
\affiliation{Dynamitron Tandem Labor des RUBION, Ruhr-Universit\"at Bochum, 44780 Bochum, Germany}
\author{J.~Endres}
\affiliation{Institute for Nuclear Physics, University of Cologne,  50937 K\"oln, Germany}
\author{J.~Mayer}
\affiliation{Institute for Nuclear Physics, University of Cologne,  50937 K\"oln, Germany}
\author{U.~Giesen}
\affiliation{Physikalisch-Technische Bundesanstalt (PTB), 38116 Braunschweig, Germany}
\author{D.~Rogalla}
\affiliation{Dynamitron Tandem Labor des RUBION, Ruhr-Universit\"at Bochum, 44780 Bochum, Germany}
\author{F.~Schl\"uter}
\altaffiliation[Present address: ]{Forschungszentrum J\"ulich, 52425 J\"ulich, Germany}
\affiliation{Institute for Nuclear Physics, University of Cologne,  50937 K\"oln, Germany}
\author{S.G.~Pickstone}
\affiliation{Institute for Nuclear Physics, University of Cologne,  50937 K\"oln, Germany}
\author{K.O.~Zell}
\affiliation{Institute for Nuclear Physics, University of Cologne, 50937 K\"oln, Germany}
\author{A.~Zilges}
\affiliation{Institute for Nuclear Physics, University of Cologne, 50937 K\"oln, Germany}

\date{\today}
             
\begin{abstract}
\setlength{\parindent}{0pt}
\textbf{Background:} Uncertainties in adopted models of particle+nucleus optical-model potentials directly influence the accuracy in the theoretical predictions of reaction rates as they are needed for reaction-network calculations in, for instance, $\gamma$-process nucleosynthesis. The improvement of the $\alpha$+nucleus optical-model potential is hampered by the lack of experimental data at astrophysically relevant energies especially for heavier nuclei.

\textbf{Purpose:} Measuring the \isotope[187]{Re}($\alpha$,n)\isotope[190]{Ir} reaction cross section at sub-Coulomb energies extends the scarce experimental data available in this mass region and helps understanding the energy dependence of the imaginary part of the $\alpha$+nucleus optical-model potential at low energies.

\textbf{Method:} Applying the activation method, after the irradiation of natural rhenium targets with $\alpha$-particle energies of 12.4 MeV to 14.1 MeV, the reaction yield and thus the reaction cross section were determined via $\gamma$-ray spectroscopy using the Cologne Clover Counting Setup and the method of $\gamma\gamma$ coincidences.

\textbf{Results:} Cross-section values at five energies close to the astrophysically relevant energy region were measured. Statistical Model calculations revealed discrepancies between the experimental values and predictions based on widely used $\alpha$+nucleus optical-model potentials. However, an excellent reproduction of the measured cross-section values could be achieved from calculations based on the so-called Sauerwein-Rauscher $\alpha$+nucleus optical-model potential.

\textbf{Conclusion:} The results obtained indicate that the energy dependence of the imaginary part of the $\alpha$+nucleus optical-model potential can be described by an exponential decrease. Successful reproductions of measured cross sections at low energies for $\alpha$-induced reactions in the mass range $141\leq A \leq 187$ confirm the global character of the Sauerwein-Rauscher potential.
\end{abstract}

\maketitle

\section{Introduction}
The $\gamma$ process \cite{woosley, rayet}  is an important part of the nucleosynthesis of the $p$ nuclei, \textit{i.e.}, those 30 to 35 neutron-deficient heavy isotopes whose solar system abundance cannot be explained by neutron-capture processes \cite{nemeth, arlandini, arnould, rauscherRep}. According to this concept, in an explosive astrophysical scenario (ccSN \cite{Rauscher2002, Woosley2007} or type Ia-SN \cite{Travaglio2011}) energetic photons photodissociate seed nuclei mainly via ($\gamma$,n) reactions due to the lower separation energy within an isotopic chain. Branching points where the $\gamma$-process path can be deflected towards other isotopic chains are determined by the probabilities of ($\gamma$,p) and ($\gamma$,$\alpha$) reactions.

For the prediction of branching points and for reliable calculations of isotopic abundances stemming from the $\gamma$-process reaction network, reaction rates of the participating reactions have to be known precisely. The enormous number of reactions as well as the fact that many reactions are not easily accessible via experimental measurements create needs for theoretical predictions of reaction cross sections. For this purpose, the Hauser-Feshbach Statistical Model is widely used \cite{Hauser1952, Rauscher2000}. The uncertainties in its predictions can be traced back to uncertainties in nuclear physics input parameters such as $\gamma$-strength functions, nuclear level densities, and particle+nucleus optical-model potentials \cite{arnould, Rauscher2011}. 

The absence of reliable experimental data, in particular for very heavy nuclei, hampers the drawing of final conclusions about the quality of, for instance,  adopted models of $\alpha$+nucleus optical-model potentials ($\alpha$-OMPs). For very heavy nuclei with masses $A > 140$, the ($\gamma$,$\alpha$) branching becomes more and more important and, therefore, also the influence of the adopted models for $\alpha$-OMPs. Usually, the parameters of an optical-model potential are fitted to scattering data far above the Coulomb barrier \cite{arnould}. But extrapolations of parametrizations of $\alpha$-OMPs to the astrophysically relevant energy region often fail to reproduce cross-section measurements for $\alpha$-capture reactions (see, \textit{e.g.}, \cite{Kiss2011a, Yalcin2009, Harissopulos2005, Gyurky2006, Gyurky2010, Rapp2002, Rapp2008, Sauerwein} and references therein). This is mainly due to the energy dependence of the imaginary part of the $\alpha$-OMP. It was found that at center-of-mass energies below the Coulomb barrier the depth of the $\alpha$-OMP should be strongly decreased \cite{Avrigeanu2010}.

Since $\alpha$-scattering becomes indistinguishable from Rutherford scattering at sub-Coulomb energies, investigations of the character of the $\alpha$-OMP can be better achieved by the study of $\alpha$-induced reaction cross sections as is done in this work.  Concerning $\gamma$-process nucleo\-synthesis, the investigation of ($\alpha$,$\gamma$) reactions would be the preferable choice. At energies achievable in experiments, ($\alpha$,$\gamma$) reaction cross sections are usually quite sensitive to the $\gamma$-width which is composed of $\gamma$-strength function and nuclear level densities which are uncertain as well \cite{arnould, Rauscher2011, rauscherRep}. Therefore, uncertainties in theoretical predictions are not only influenced by the adopted $\alpha$-OMPs but also by the uncertainties of the input of the $\gamma$-width.

\begin{figure}[t]
\centering
\includegraphics[width=0.5\textwidth]{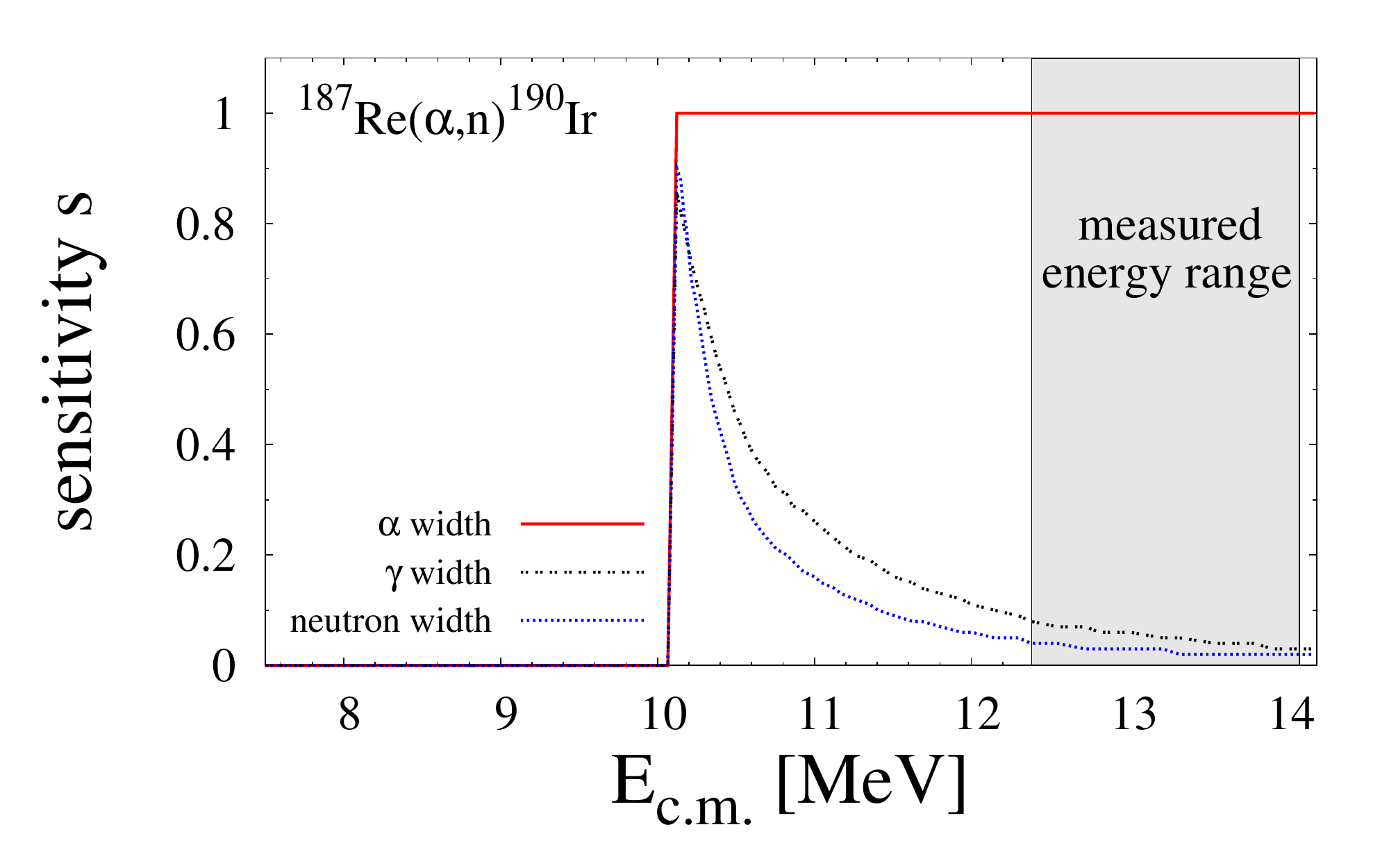}
\caption{(Color online) Sensitivity of the laboratory cross section for the  \isotope[187]{Re}($\alpha$,n)\isotope[190]{Ir} reaction to the variation of the $\alpha$ (red line), $\gamma$ (black dotted line), and neutron widths (blue dashed line) as a function of center-of-mass energy \cite{sensi}.
 The cross section is insensitive to the proton width. See text for details.}
\label{fig:sensi}
\end{figure}

For ($\alpha$,n) reactions, this is often not the case. Figure \ref{fig:sensi} shows the sensitivity of the \isotope[187]{Re}($\alpha$,n)\isotope[190]{Ir} reaction cross section for the different decay widths as a function of center-of-mass energy. 

The sensitivity is defined as  the relative variation of the cross section $\frac{\sigma'}{\sigma}$ by varying the respective decay width by a factor $f=\frac{\Gamma'}{\Gamma}$ \cite{sensi}:
\begin{align}
s=\frac{\frac{\sigma'}{\sigma}-1}{f-1}.
\end{align}
Above the ($\alpha$,n) reaction threshold, the sensitivity of the cross section to the $\gamma$ and neutron widths decreases exponentially until the reaction probability is exclusively sensitive to the $\alpha$ width and, therefore, to the $\alpha$-OMP. In this work, the \isotope[187]{Re}($\alpha$,n) reaction cross section was measured at five $\alpha$-energies between 12.4 MeV and 14.1 MeV, and thus in an energy region where the reaction is an excellent candidate to study the $\alpha$-OMP.

The cross section was measured by applying the activation technique. This technique has been widely used before and will be introduced only briefly in Sec. \ref{sec:activationTechnique}. Details about the target preparation and the activation at the Physikalisch-Technische Bundesanstalt (PTB) in Braunschweig, Germany will be given in Sec. \ref{sec:target}. The produced \isotope[190]{Ir} nuclei decay with a half-life of $T_{1/2} = 11.8(1)$ d via electron-capture decay ($\epsilon$-decay) to \isotope[190]{Os} \cite{NNDC}. By means of high-resolution $\gamma$-ray spectroscopy, the number of de-exciting \isotope[190]{Os} nuclei can be measured and hence the number of \isotope[190]{Ir} nuclei  produced can be calculated. In the present case, the $\gamma$-ray spectroscopy was performed using the Cologne Clover Counting Setup. This experimental setup was specifically designed for nuclear astrophysics purposes and will be described in Sec. \ref{sec:CologneClover}. The data analysis, the resulting cross-section values and comparisons to Statistical Model calculations will be presented in Sec. \ref{sec:analysis} and \ref{sec:results}.

\section{The Activation Technique}
\label{sec:activationTechnique}
During the irradiation the number of produced \isotope[190]{Ir} nuclei $N_{\text{Ir}}$ is given by
\begin{align}
\dot{N}_{\text{Ir}} = \sigma \Phi(t) N_{\text{Re}} - \lambda N_{\text{Ir}},
\end{align}
where $\Phi(t)$ denotes the flux of $\alpha$-particles, $\sigma$ the cross section, $N_{\text{Re}}$ the number of \isotope[187]{Re} nuclei within the target material, and $\lambda$ the decay constant of \isotope[190]{Ir}. For a constant flux of projectiles ($\Phi(t)=\Phi$) and a vanishing number of \isotope[190]{Ir} nuclei at the beginning of the activation ($N_{\text{Ir}}(t_0)=0$) the solution of this differential equation is given by
\begin{align}
\label{eq:solutionDGL}
N_{\text{Ir}}(t)=\frac{\sigma \Phi N_{\text{Re}}}{\lambda} \left(1-e^{-\lambda t}\right).
\end{align}
Generally, the flux of projectiles is not constant over the whole irradiation period due to several technical limitations, $e.g.$, extraction of ions out of the ion source, transmission through the accelerator, etc. However, this problem can be solved by measuring the accumulated charge on the target during the activation in short time intervals. Assuming that the flux of projectiles is constant within these short time intervals, Eq. \ref{eq:solutionDGL} can be solved iteratively for the whole irradiation time:
\begin{align}
\label{eq:iterative}
N_{\text{Ir}}^{i+1}= N_{\text{Ir}}^i e^{-\lambda \Delta t_{i+1}} + \frac{\sigma \Phi^{i+1} N_{\text{Re}}}{\lambda} \left(1-e^{-\lambda \Delta t_{i+1}} \right).
\end{align}
After the irradiation, the number of \isotope[190]{Ir} nuclei can be determined by the spectroscopy of their decay. In the present case, this was done by observing the consecutive $\gamma$-decay of levels populated in \isotope[190]{Os} using the Cologne Clover Counting Setup (see Sec. \ref{sec:CologneClover}). By taking into account the number of events in the full-energy peak $N_{\gamma}$, the absolute full-energy peak efficiency $\epsilon_{\gamma}$ and the relative live time $\tau$ of the experimental setup as well as the absolute intensities of the specific $\gamma$-ray transition $I_{\gamma}$, the cross section of the reaction can be calculated via

\begin{align}
\label{eq:crossSection}
\sigma = \frac{\lambda \cdot N_{\gamma} \cdot e^{\lambda t_W}}{N_{\text{Re}} \tau I_{\gamma} \epsilon_{\gamma} \left(1-e^{-\lambda t_C}\right)} \times f_{ac}^{-1} .
\end{align}

Here, $t_W$ denotes the time between the end of the irradiation and the beginning of the counting and $t_C$ the time of measurement with the Cologne Clover Counting Setup, respectively. The factor $f_{ac}$ corrects for decaying nuclei during the irradiation and the non-constant particle-flux 

\begin{align}
f_{ac} = \sum \Phi^i \cdot  \left(1-e^{-\lambda \Delta t_i}\right) \cdot e^{-\lambda (t_A-t_i)}.
\end{align}

In this formula, $\Phi_i$ denotes the $\alpha$-particle flux in a time interval $\Delta t_i$, $t_i$ the sum of the first $i$ time intervals, and $t_A$ the end of the activation.

\section{Target Preparation and Activation}
\label{sec:target}
Rhenium targets with natural isotopic abundance were used, consisting of 62.6 \% \isotope[187]{Re} and 37.4 \% \isotope[185]{Re}. Aluminum was used as backing material. To avoid chemical reactions of rhenium with aluminum, an additional layer of tantalum was placed between the target and the backing material. 

The thicknesses of the targets were determined by Rutherford Backscattering Spectrometry (RBS) at the RUBION Dynamitron Tandem accelerator in Bochum, Germany. This method is based on the elastic scattering of ions, in this case He$^+$ ions, in the target material under backward angles. The thickness of the target can then be calculated from the energy loss of the ions inside the target material. The initial energy of the He$^+$ ions was $2$ MeV $\pm$ $1$ keV. For the detection of the ions, a silicon detector was placed at a distance of 35 mm from the center of the target resulting in a solid angle coverage of $1.91 \pm 0.07$ msr.  For all five targets, the measured thicknesses were between 135 $\mu$g/cm$^2$ and 212 $\mu$g/cm$^2$. To ensure no material loss during the activation and the transportation, an additional RBS measurement was performed afterwards. It was shown that within the uncertainties, no material vanished.

The targets were activated at five different $\alpha$-particle energies between 12.4 MeV and 14.1 MeV at the cyclotron of PTB in Braunschweig, Germany \cite{brede}. The activation durations were between 3 h and 24 h at $\alpha$-beam currents between 1.0 $\mu$A and 2.1 $\mu$A. The energy of the $\alpha$-particles was determined by two analyzing magnets and a time-of-flight measurement with an uncertainty of $\pm$ 25 keV \cite{boettge}. An additional uncertainty in the activation energy due to the thickness of the targets was considered. The effective $\alpha$-particle energy $E_{\alpha}$ was obtained by correcting the incident energy of the $\alpha$-particles $E_0$ by their energy loss $\Delta E$:
 \begin{align}
 E_{\alpha} = E_0 - \frac{\Delta E}{2}.
 \end{align}
The energy loss was obtained using TRIM (v. 2013) simulations  \cite{ziegler}. Depending on the thicknesses of the targets and the incident $\alpha$-particle energies, the energy loss was found to be between 18 keV and 32 keV. Additional calculations using LISE++ \cite{lise++} were in excellent agreement with the TRIM simulations. 
During the activation, the targets were water-cooled to guarantee the thermodynamic stability of the target material. A cooling-trap at liquid-nitrogen temperature was installed to avoid build up of carbon on the target surface. The $\alpha$-particles were stopped in the aluminum backing and the target holder served as a Faraday cup. For a reliable charge collection, a negative voltage of $U_S = -300$V was applied to the aperture of the chamber to suppress secondary electrons. The $\alpha$-beam current was measured by integrating the collected charge on the target in intervals of 60 s. For the charge collection an uncertainty of 1\% was taken into account. 
Detailed information about the targets and activation parameters is listed in Table \ref{tab:parameters}.

\begin{table}
\centering
\caption{$\alpha$-particle energies $E_{\alpha}$ , target thickness $d$, beam currents $\Phi_{\alpha}$, time of irradiation $t_A$, and counting $t_C$ for each target. For an $\alpha$-particle energy of 12.4 MeV the target was activated within two runs.}
\label{tab:parameters}
\vspace{2mm}
\renewcommand{\baselinestretch}{1.5}\normalsize
\begin{tabular}{cScrr}
\hline
\hline
$E_{\alpha}$ [keV] & {$d$ [$\mu$g/cm$^2$]} &{$\Phi_{\alpha}$ [$\mu$A]}&$t_A$ [h]&$t_C$ [h] \\
\colrule
14091 $\pm$ 28 &136(13) &1.4-1.9&3.2&508\\
13689 $\pm$ 28& 153(11) &1.6-1.8&6.8&202\\
13286 $\pm$ 29& 202(7) &1.0-2.0&7.3&334\\
12785 $\pm$ 29&205(14) &1.9-2.1&23.9&110\\
12384 $\pm$ 29&212(15) &1.0-1.9&20.4 and 23.4&69\\							
\hline
\hline
\end{tabular}

\end{table}

\section{$\gamma$-ray spectroscopy at the Cologne Clover Counting Setup}
\label{sec:CologneClover}
\begin{figure}[t]
\centering
\includegraphics[width=0.50\textwidth]{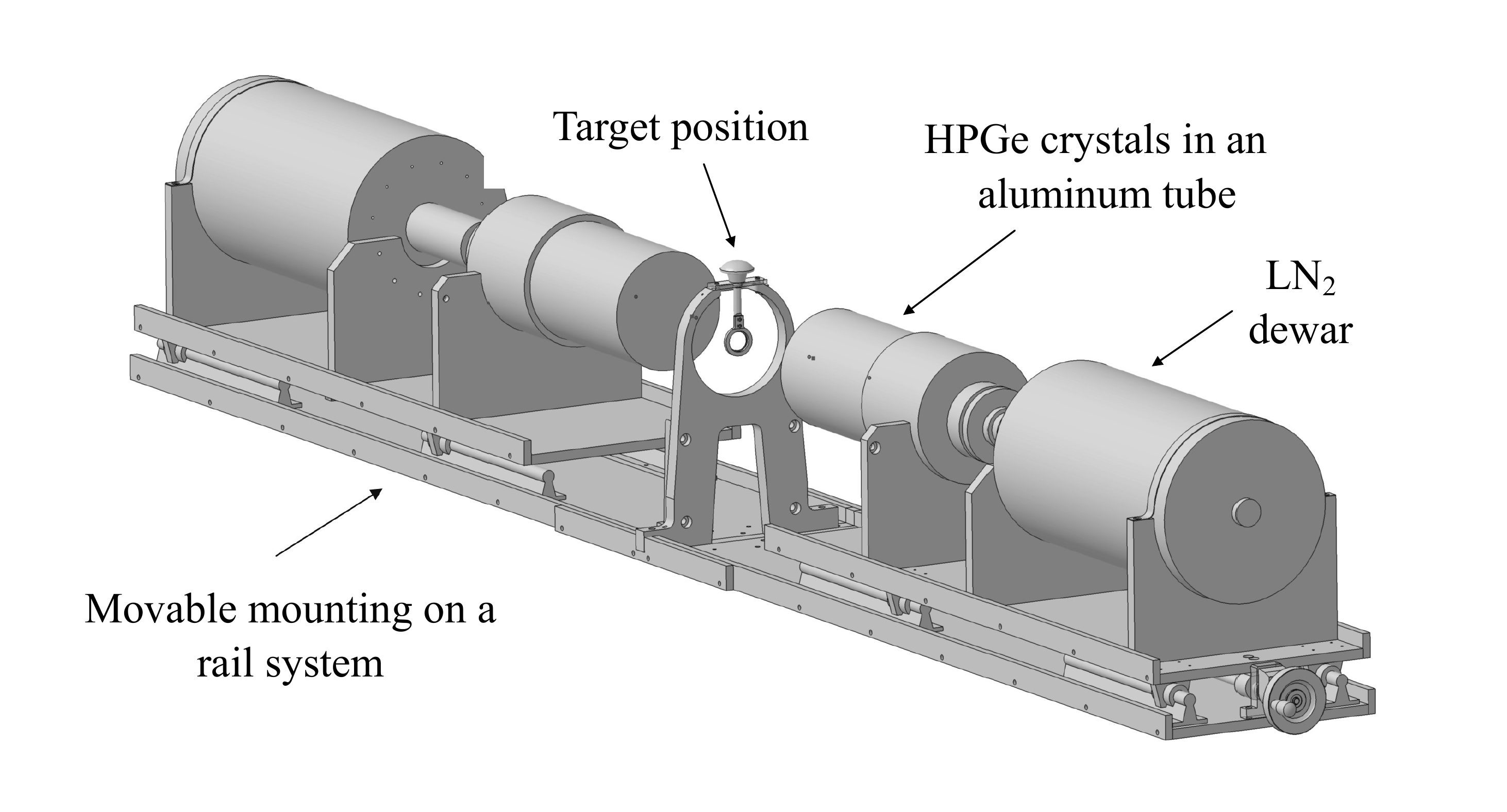}
\caption{A CAD drawing of the Cologne Clover Counting Setup. Two high-purity clover-type germanium detectors are mounted on a movable rail system. The target is placed between them. In the closest geometry, the detectors cover a solid angle of almost 4$\pi$.}
\label{fig:clover_setup}
\end{figure}

After the irradiation, the targets were transported to Cologne to be analyzed using the Cologne Clover Counting Setup at the Institute for Nuclear Physics of the University of Cologne. This setup has successfully been used before to determine cross sections of $\alpha$-induced reactions, see, $e.g.$, \cite{Sauerwein, Netterdon}. 

The Cologne Clover Counting Setup consists of two clover-type high-purity germanium (HPGe) detectors in a close face-to-face geometry covering a solid angle of almost 4$\pi$. Due to the smaller crystals in comparison to a single-crystal HPGe detector of the same size, summing effects are less pronounced. Furthermore, the higher granularity allows the measurement of $\gamma\gamma$ coincidences between the clover leaves. As it can be seen in Fig. \ref{fig:clover_setup}, both Clover detectors are mounted on a rail system allowing a precise variation of the distance between the target and the detectors. In addition, both Clover detectors can be equipped with BGO shields for an active Compton background suppression. The whole setup is shielded by 10 cm of lead and 3 mm of copper against natural radioactivity and X-ray radiation, respectively.

The preamplifier signals of each Clover leaf are processed digitally using DGF-4C Rev. F modules from the company XIA \cite{xia1, xia2}. The digitization is performed with a depth of 14 bit and a frequency of 80 MHz. After the extraction of energy and time information, the data is stored in an event-by-event listmode format. This allows the offline analysis of $\gamma\gamma$ coincidences which is important in order to further reduce the background in the measured spectra and, therefore, to determine the total reaction cross section in a range of a few $\mu$b (see Sec. \ref{sec:gammagamma}). 

The efficiency calibration for the setup was performed using \isotope[152]{Eu}, \isotope[57]{Co}, and \isotope[137]{Cs} calibration sources. For a reliable full-energy peak efficiency, corrections considering summing effects have been taken into account. For this, two measurements were performed. One at a target-detector distance of $10$ cm where summing effects are negligible and one at a distance of $1.3$ cm. The photopeak efficiencies of $\gamma$-ray transitions of \isotope[57]{Co} and \isotope[137]{Cs} which are not affected by coincidence summing were used to calculate a scaling factor which accounts for the difference in geometrical efficiency between the far and the close distance measurement. The measured values of \isotope[152]{Eu} were fitted using a sum of exponential functions. This fit function was then rescaled using the geometrical scaling factor to obtain a function which represents the photo-peak efficiency of the Cologne Clover Counting Setup in the close distance geometry.

\begin{figure}[t]
\centering
\includegraphics[width=0.49\textwidth]{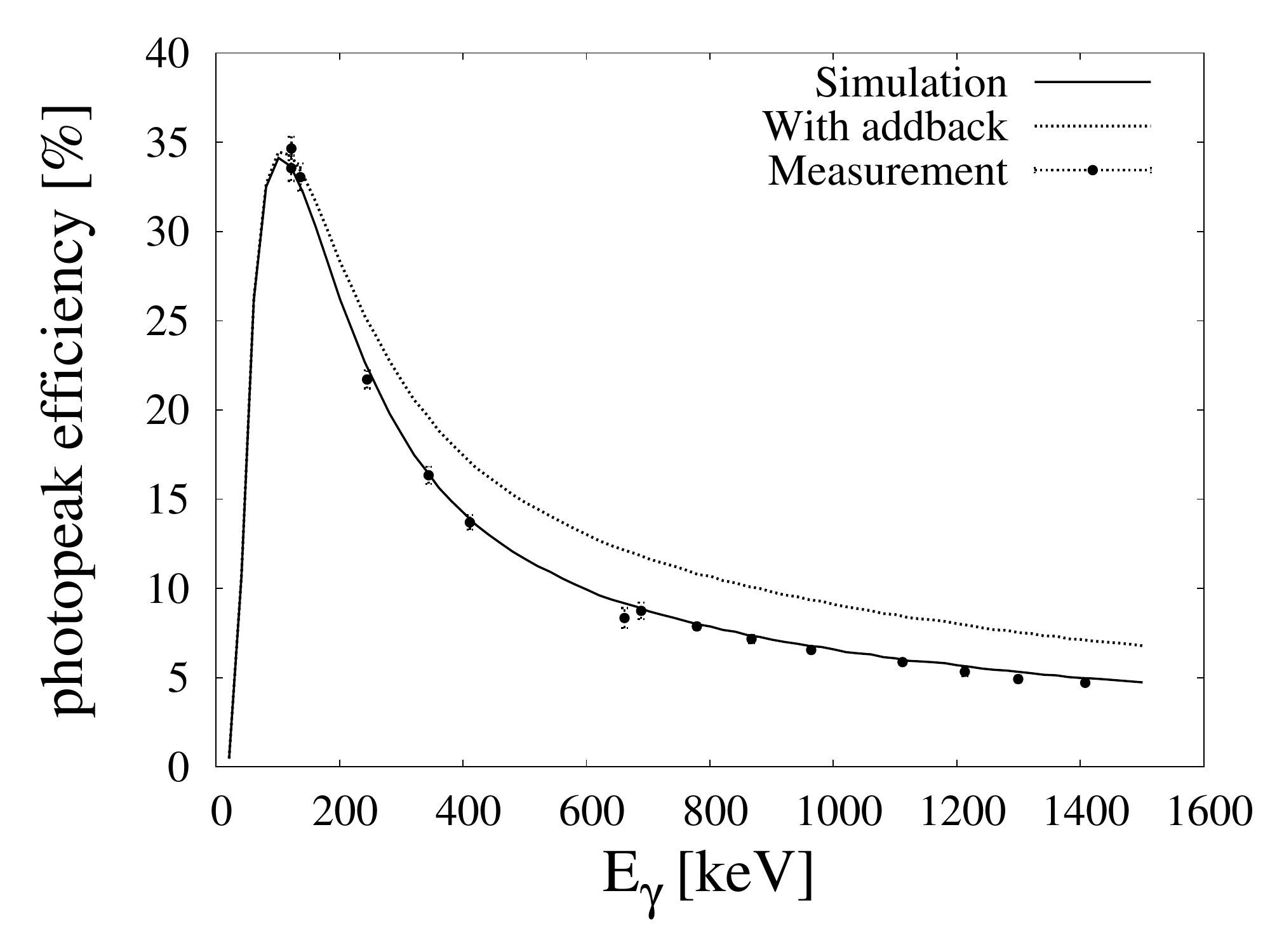}
\caption{Total full-energy peak efficiency of the Cologne Clover Counting Setup as a function of $\gamma$-ray energy. Additionally, a simulation of the full-energy peak efficiency obtained with Geant4 with and without addback is shown. In favor of a better readability, the fitted efficiency curve is not shown.}
\label{fig:efficiency}
\end{figure}

Additionally, the Cologne Clover Counting Setup was implemented in a Geant4-based Monte Carlo simulation \cite{geant4-1, geant4-2}. Hence, it was possible to obtain a simulated full-energy peak efficiency in the energy region between 20 keV and 1500 keV. In Fig. \ref{fig:efficiency} the measured full-energy peak efficiencies as well as the simulated efficiency curve are presented as a function of the $\gamma$-ray energy. 

Due to the granularity of the Clover detectors, the geometrical efficiency is reduced in comparison to a one-crystal HPGe detector of the same composite size resulting in a smaller efficiency for a complete energy deposition. An incomplete energy deposition of $\gamma$-rays leads to an increased Compton background and a decreased full-energy peak efficiency \cite{Duchene}. However, Compton-scattered $\gamma$-rays could enter and deposit their remaining energy in the active volume of another Clover leaf. Adding back the energy of coincident events of one Clover detector could thus lead to a reproduction of the full-energy event. This so-called addback increases the efficiency of the detection system, although it leads to a worse energy resolution \cite{Duchene}.

To study the influence of an addback on the efficiency of the detection setup, an algorithm for the summing procedure was implemented in the analysis software and tested performing calibration measurements with \isotope[152]{Eu} and \isotope[60]{Co}. The distance between the source and the detectors during this measurement was equal to the one used for the activation measurement. 

The algorithm provides three different operation modes for the Clover detectors. In singles mode, all crystals of the Clover are treated as individual detectors and the resulting spectrum represents the sum of all singles spectra. In direct mode, all events with multiplicity $m > 1$ are ignored. The addback-mode spectrum is obtained by the sum of the direct-mode spectrum and the reconstructed events by applying the addback.

\begin{figure}[t]
\centering
\includegraphics[width=0.49\textwidth]{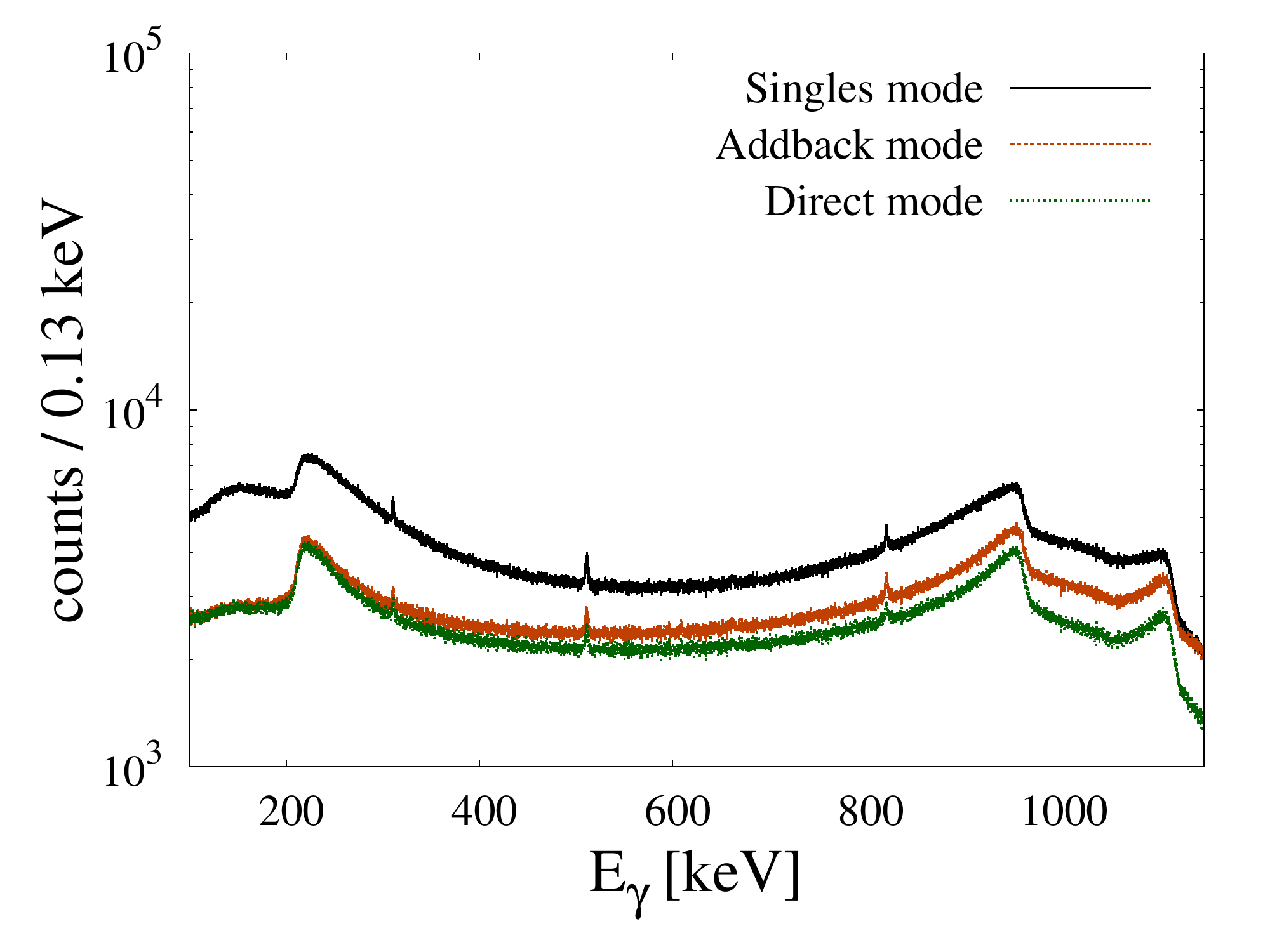}
\caption{(Color online) Comparison of Compton background of the different Clover operation modes. See text for details.}
\label{fig:comton}
\end{figure}

The gain in efficiency by applying an addback can be expressed by a so-called addition factor \cite{Duchene}:
\begin{align}
\label{eq:addFac}
g=\frac{\epsilon_{add}}{\epsilon_{dir}}
\end{align}
This formula expresses the ratio between the number of events in the full-energy peak of the spectrum obtained by operating the Clover Setup in addback and direct modes. This addition factor $g$ was found to depend logarithmically on the $\gamma$-ray energy \cite{Duchene}. For the Cologne Clover Counting Setup, the gain in efficiency can be described by:
\begin{align}
\label{eq:addFunc}
g(E_{\gamma})=0.19(2) \times \log{E_{\gamma}}-0.92(1)
\end{align}
The photopeak efficiency of the Cologne Clover Counting Setup operated in addback mode is also illustrated in Fig. \ref{fig:efficiency}. Figure \ref{fig:comton} shows a comparison of the Compton background of $\gamma$-ray spectra measured in direct, singles, and addback modes for a \isotope[60]{Co} calibration source. The comparison shows the expected strong reduction of the Compton background by applying an addback to the measured data. The spectra were taken for the same distance between the source and the detectors as it was during the activation experiment. The average counting rate in the detectors was about 3 kcps. The relative reduction of all events in the Compton background for the addback-mode spectrum was about 33 \% for the measurement of the \isotope[60]{Co} calibration source in comparison to the singles-mode spectrum.

\begin{figure}[t]
\centering
\includegraphics[width=0.49\textwidth]{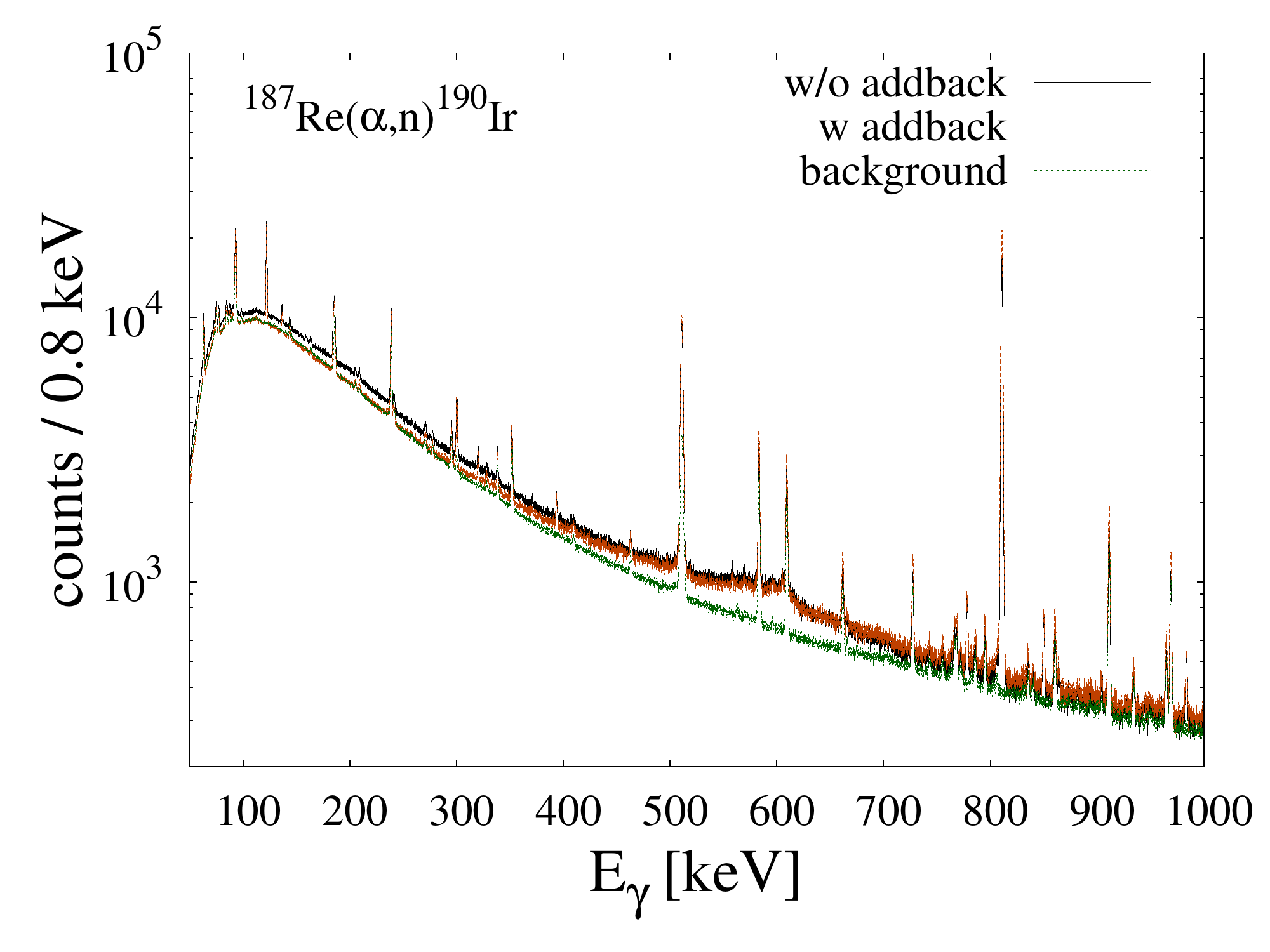}
\caption{(Color online) Example of the measured $\gamma$-ray spectra for an alpha-particle energy of 13.7 MeV. For comparison, the spectrum obtained by the application of an addback as well as the measured background spectrum are shown in addition.}
\label{fig:addback1}
\end{figure}

Applying the addback to the $\gamma$-ray spectroscopy data of the activated \isotope[187]{Re} targets does not lead to satisfactory results. Due to the $\gamma$-ray transition cascades in \isotope[190]{Os} and the increased geometrical efficiency by operating the Clovers in addback mode, summing-effects are more pronounced, leading to a reduction in photopeak efficiency.
In Fig. \ref{fig:addback1} the $\gamma$-ray spectra of \isotope[190]{Os} are shown with and without the application of the addback. The measured background spectrum was rescaled and also drawn for comparison.
In contrast to the measurement of the \isotope[60]{Co} calibration source, see Fig. \ref{fig:comton}, by applying an addback to the data of the activation experiment, the reduction of the Compton background is strongly decreased. This indicates already that the main source of the measured background stems from Compton scattered events from the environment which have deposited their energy partly in the crystals of the Clovers and partly in environmental material. Since such events cannot be reconstructed by applying an addback, the Compton-background reduction is smaller than it is needed to balance the more pronounced summing-effects, especially in the relevant energy region between 360~keV and 615 keV (see Fig. \ref{fig:addback1}). Thus, the usage of an addback leads to peak-to-background ratios which are smaller than those in the summed singles spectra and was therefore not used for the further analysis.

\section{Data Analysis}
\label{sec:analysis}
\subsection{$\gamma$-decays of \isotope[190]{Os} levels}
\begin{figure}[t]
\centering
\includegraphics[width=0.35\textwidth]{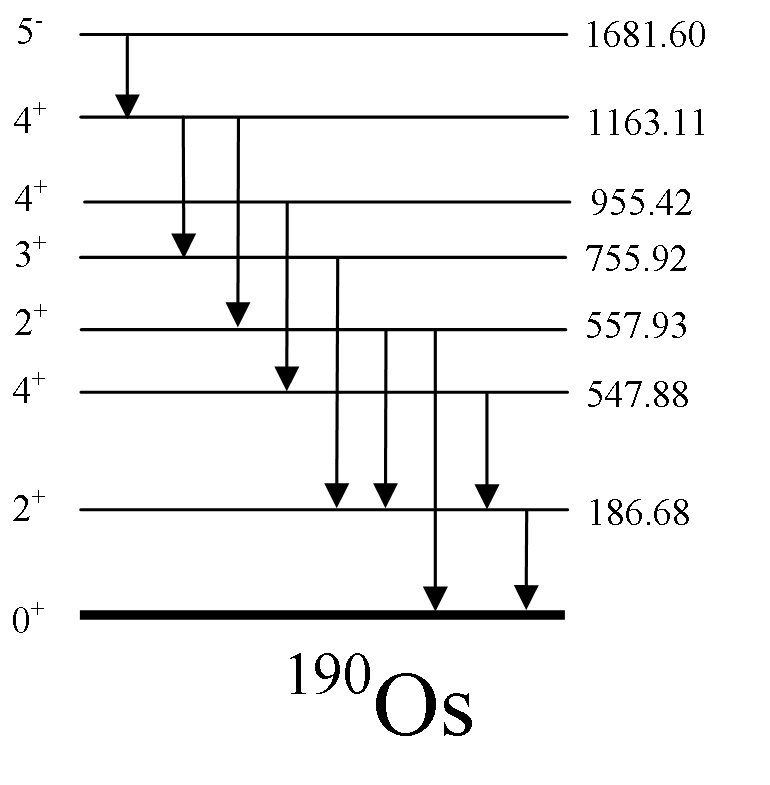}
\caption{Simplified level scheme of \isotope[190]{Os}. Only the strongest $\gamma$-ray transitions are shown in the level scheme. All data are adopted values from \cite{NNDC}. Note that there are two 407 keV $\gamma$-ray transitions contributing to the specific peak in the $\gamma$-ray spectra. Besides the 557.9 keV $\gamma$-ray transition, all visible transitions are in coincidence with the $2^+ \rightarrow 0^+$ ground-state transition with a $\gamma$-ray energy of 186.6 keV. The parameters of the $\gamma$-ray transitions are given in Table \ref{tab:gammas}. }
\label{fig:levels}
\end{figure}

Figure \ref{fig:levels} shows a simplified level scheme of the \isotope[190]{Os} nucleus where only the strongest $\gamma$-ray transitions are shown. For an activation energy of $E_{\alpha} = 13.7$ MeV, seven $\gamma$-ray transitions of \isotope[190]{Os} could be identified. Unfortunately, the strongest $\gamma$-ray transition with a $\gamma$-ray energy of 186.68 keV could not be used for the determination of the total cross section due to contributions stemming from contaminants in the target material activated in the beam.

Although using clover-type HPGe detectors should reduce summing effects, the summing-out of events is still not negligible. 

For this, the Geant4-based Monte Carlo simulation was used in order to study the summing effects due to the observed $\gamma$-ray cascades, see Fig. \ref{fig:levels}. The full-energy peak efficiency of all $\gamma$-ray transitions was simulated twice. First, as a single $\gamma$-ray transition and afterwards within its specific $\gamma$-ray cascade. The ratio of the full-energy peak efficiencies yields the correction factor for summing out. Hence, it is possible to correct the obtained events in the full-energy peak for summing effects. The contribution of summed-out events was found to vary between 6.9\% and 22.2\% depending on the specific $\gamma$-ray transition.
Table \ref{tab:gammas} lists the strongest $\gamma$-ray transitions in \isotope[190]{Os}, their absolute $\gamma$-ray intensities $I_{\gamma}$, their photopeak efficiencies $\epsilon_{\gamma}$ and their summing correction factor $\xi$.

In the $\gamma$-ray spectra with the highest intensities of \isotope[190]{Os}, the peak-to-background ratios were found to be between 0.5 \% and 4 \%. Therefore, for the analysis of the $\gamma$-ray spectra of the other targets a strong background reduction had to be applied for a reliable determination of the events in the full-energy peaks.

\begin{table}
\caption{$\gamma$-ray energies $E_{\gamma}$, absolute $\gamma$-ray intensities $I_{\gamma}$ and photopeak efficiencies $\epsilon_{\gamma}$ as well as summing-correction factors $\xi$ and $\eta$-parameters for the $\gamma$-ray transitions. The $\gamma$-ray energies $E_{\gamma}$ and the absolute $\gamma$-ray intensities $I_{\gamma}$ are adopted values of Ref. \cite{NNDC}. The values of the photopeak efficiencies were calculated using the fitted efficiency function as described in Sec. \ref{sec:CologneClover} and the summing correction factors were obtained via simulation of the specific $\gamma$-ray cascades using the Geant4-based simulation. Additionally, the $\eta$ parameters calculated via Eq. \ref{eq:eta} for the gate on the 186.68 keV ground-state transition are listed. See text for details.}
\label{tab:gammas}
\vspace{2mm}
\renewcommand{\baselinestretch}{1.5}\normalsize
\begin{tabular}{lSSSS}
\hline
\hline
$E_{\gamma}$ [keV] & {$I_{\gamma}$} [\%] & {$\epsilon_{\gamma}$ [\%]} & {$\xi$ [\%]} & {$\eta$}\\
\hline
371.24 & 22.8(7) & 15.1(12) & 12.0(5) & 5.7(7)\\
407.22 & 28.5(8) & 13.9(10) & 18.9(15)& 7.1(8)\\
518.55 & 34.0(15) & 11.1(9) & 18.9(7) &    \\
557.95 & 30.1(13) & 10.4(12) & 10.6(5) &  \\
569.30 & 28.5(13) & 10.2(9) & 6.9(4) &  \\
605.14 & 39.9(18) & 9.7(9) & 22.2(8) & 14.4(28)\\
\hline
\hline
\end{tabular}
\renewcommand{\baselinestretch}{1.0}\normalsize
\end{table}

\subsection{$\gamma\gamma$-coincidence method}
\label{sec:gammagamma}
\begin{figure}[t]
\centering
\includegraphics[width=0.49\textwidth]{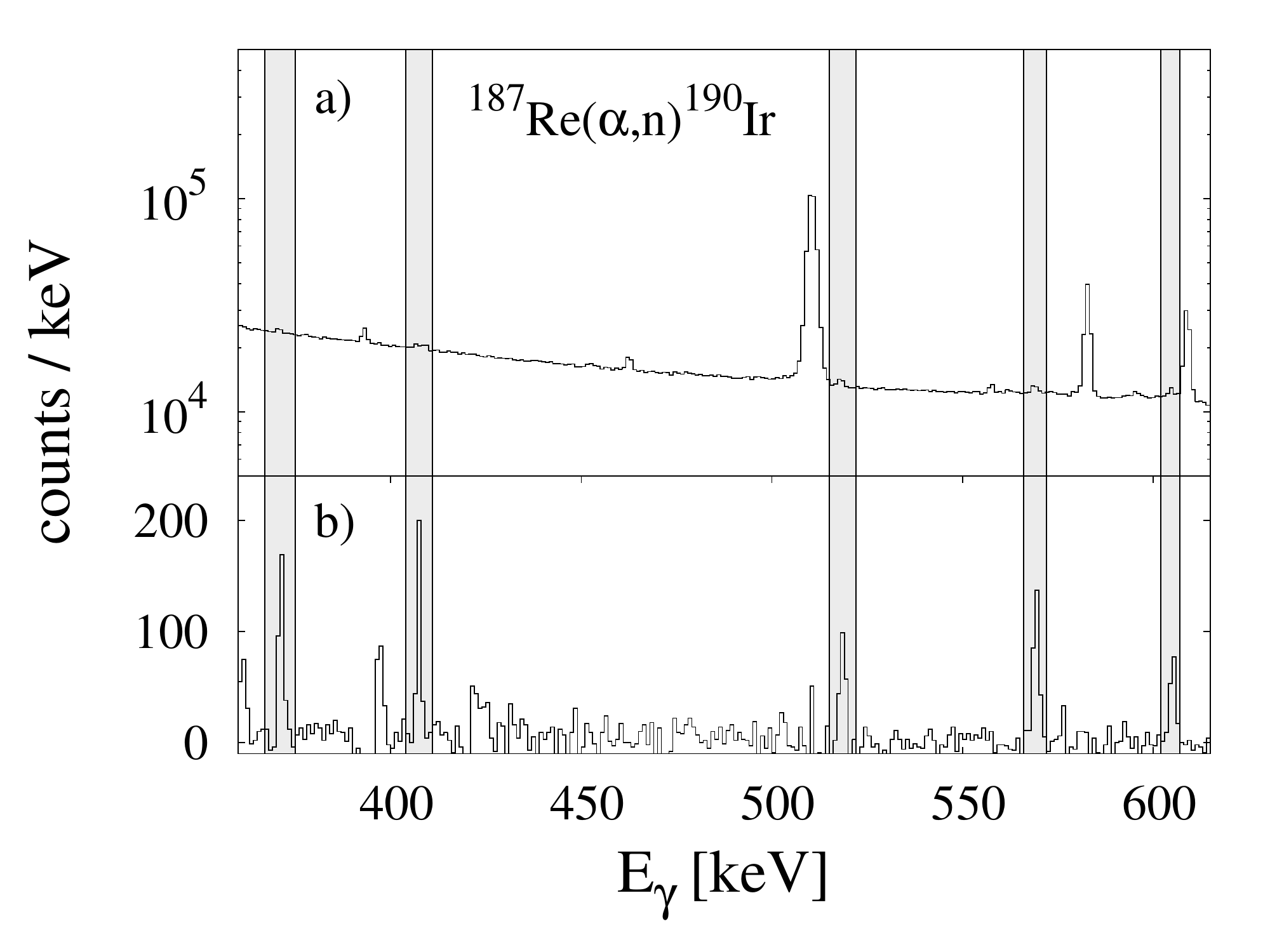}
\caption{Illustration of the $\gamma\gamma$-coincidence method. The upper panel shows the sum of the singles spectra of all clover leaves in an energy range between 360 keV and 615 keV. The measuring time was 202 h. The spectrum in the lower panel shows only those events which were in coincidence with an event with an energy of 186.68 keV. The peak-to-background ratio is increased by a factor of 2 to 10 depending on the specific $\gamma$-ray transition. See text for details.}
\label{fig:vergleich}
\end{figure}

Since the activity of the activated targets was low, the $\gamma$-ray intensities in the measured $\gamma$-ray spectra were not sufficient for all $\alpha$-particle energies. Therefore, the $\gamma\gamma$-coincidence method was used for further reducing the background in the measured $\gamma$-ray spectra.  This method is based on the selective observation of detected $\gamma$-ray events which are in coincidence with another event with an energy in a specific range. In Fig. \ref{fig:vergleich}, the power of the $\gamma\gamma$-coincidence method is illustrated for the case of the \isotope[187]{Re}($\alpha$,n)\isotope[190]{Ir} reaction at an energy of $E_{\alpha}$=13.7 MeV. Figure \ref{fig:vergleich} a) shows the summed singles spectra of both Clover detectors for an energy region between 360 keV and 615 keV. Five of the most intense $\gamma$-ray transitions in \isotope[190]{Os} are in coincidence with the 186.6 keV ground-state transition, see Fig. \ref{fig:levels}. By taking into account only those events observed in coincidence with this ground-state transition, the peak-to-background ratio can be improved by a factor of 2 to 10, see Fig. \ref{fig:vergleich} b). This enables a reliable determination of the number of events in the full-energy peaks. For the determination of total cross-section values a precise knowledge of the full-energy peak efficiency is of utmost importance. The number of nuclei $D(t_1,t_2)$ decaying in a time period $t_2-t_1$ can be derived from
\begin{align}
D(t_1,t_2)=\frac{N_{\gamma}(E_{\gamma})}{ \epsilon_{\gamma} I_{\gamma} \tau}
\end{align}
where $N_{\gamma}$ is the number of events observed in the full-energy peak in the $\gamma$-ray spectrum for a $\gamma$-ray energy of $E_{\gamma}$, $ \epsilon_{\gamma}$ the total full-energy peak efficiency, $\tau$ the correction for dead-time effects, and $I_{\gamma}$ the absolute $\gamma$-ray intensity. The number of events $N_{\gamma}(E^{cut}_{\gamma},E_{\gamma})$ observed in the full-energy peak in the coincidence spectrum obtained by  gating on events with an energy of $E^{cut}_{\gamma}$ is 
\begin{align}
 N_{\gamma}(E^{cut}_{\gamma},E_{\gamma}) \propto D(t_1,t_2) \cdot \epsilon_{\gamma} I_{\gamma} \epsilon^{cut}_{\gamma} I^{cut}_{\gamma}.
\end{align}
The full-energy peak efficiency in the coincidence spectra can then be obtained by the product of the constant
\begin{align}
\label{eq:eta}
\eta = \frac{N_{\gamma}(E_{\gamma})}{N_{\gamma}(E^{cut}_{\gamma},E_{\gamma})}
\end{align}
and the number of events in the full-energy peak in the coincidence spectrum. These parameters are independent of the activity of the targets and depend only on the efficiencies and the absolute intensities of the observed $\gamma$-ray transitions. Consequently, these measured $\eta$-parameters can be used for the determination of the cross-section values for all the other activation energies. This method was already successfully applied during the measurement of the \isotope[141]{Pr}($\alpha$,n)\isotope[144]{Pm} cross section \cite{Sauerwein}.

\subsection{Determining total cross-section values}
The total cross-section values for the \isotope[187]{Re}($\alpha$,n)\isotope[190]{Ir} reaction were calculated according to Eq. \ref{eq:crossSection} in Sec. \ref{sec:activationTechnique}. For an $\alpha$-particle energy of 13.7 MeV, cross-section values were obtained from full-energy peak events of six $\gamma$-ray transitions. The values for the other activation energies were obtained by applying the $\gamma\gamma$-coincidence method as described in Sec. \ref{sec:gammagamma}. 

The $\eta$ parameters were first calculated for an $\alpha$-particle energy of 13.7 MeV. Only three of six parameters, $e.g.$ for the 371 keV, the 407 keV, and the 605 keV $\gamma$-ray transistion, could also be calculated from the $\gamma$-ray spectrum for an activation energy of 14.1 MeV. For the further analysis, the weighted mean of the $\eta$ parameters of these three $\gamma$-ray transitions were used to calculate the total number of full-energy peak events for the targets activated with lower $\alpha$-particle energies. 

Summing-correction factors calculated using the Geant4-based simulation were used to correct the total numbers of events in order to take summing effects into account. 
The final values for the total cross section were obtained by calculating the weighted mean of the cross sections obtained from each $\gamma$-ray transition. For the weights the statistical uncertainties for the efficiency ($<$~10~\%), the $\gamma$-ray intensities ($<$~5~\%), the $\eta$-parameters ($<$~20~\%), the summing correction ($<$~8~\%), and the number of events in the area of the full-energy peak ($<$~30~\%) were used. 

\section{Results}
\label{sec:results}
\begin{table}
\centering
\caption{Results of the activation experiment. For each $\alpha$-particle energy, the measured cross-section values for the different $\gamma$-ray transitions are shown. The last column shows the weighted average of the obtained cross-section values.}
\label{tab:results}
\vspace{2mm}
\renewcommand{\baselinestretch}{1.5}\normalsize
\begin{ruledtabular}
\begin{tabular}{cccc}

$E_{\alpha}$ [keV] & $E_{\gamma}$ [keV] &  $\sigma$ [$\mu$b] & $\bar{\sigma}$ [$\mu$b]\\
\colrule
&605& 48.4 $\pm$ 17.4 \qquad	& \\
14091 $\pm$ 28 &407& 35.3 $\pm$ 10.5 \qquad	& 30.4 $\pm$ 8.8\\
&371& 25.4 $\pm$ 7.4	\qquad & \\
\\
&605& 11.1 $\pm$ 2.4 \qquad &  \\
&569& 11.3 $\pm$ 3.1 \qquad &  \\
13689 $\pm$ 28 & 558 & 12.4 $\pm$ 3.2 \qquad &  9.9 $\pm$ 2.0\\
&518& 9.6 $\pm$ 2.6 \qquad &  \\
&407& 8.2 $\pm$ 2.4 \qquad &  \\
&371& 7.4 $\pm$ 2.5 \qquad &  \\
\\
&605& 6.9 $\pm$ 2.0 \qquad & \\
13286 $\pm$ 29&407& 4.3 $\pm$ 1.7 \qquad & 4.97 $\pm$ 1.12\\
&371& 4.4 $\pm$ 1.4 \qquad & \\
\\
&605& 1.68 $\pm$ 0.71 \qquad & \\
12785 $\pm$ 29&407& 1.10 $\pm$ 0.30 \qquad & 1.13 $\pm$ 0.29\\
&371& 1.03 $\pm$ 0.40 \qquad & \\
\\
12384 $\pm$ 29&605&0.68 $\pm$ 0.21 \qquad & 0.85 $\pm$ 0.19 \\							
&407&1.27 $\pm$ 0.35 \qquad &\\		
\end{tabular}
\end{ruledtabular}
\end{table}

\begin{figure}[t]
\centering
\includegraphics[width=0.51\textwidth]{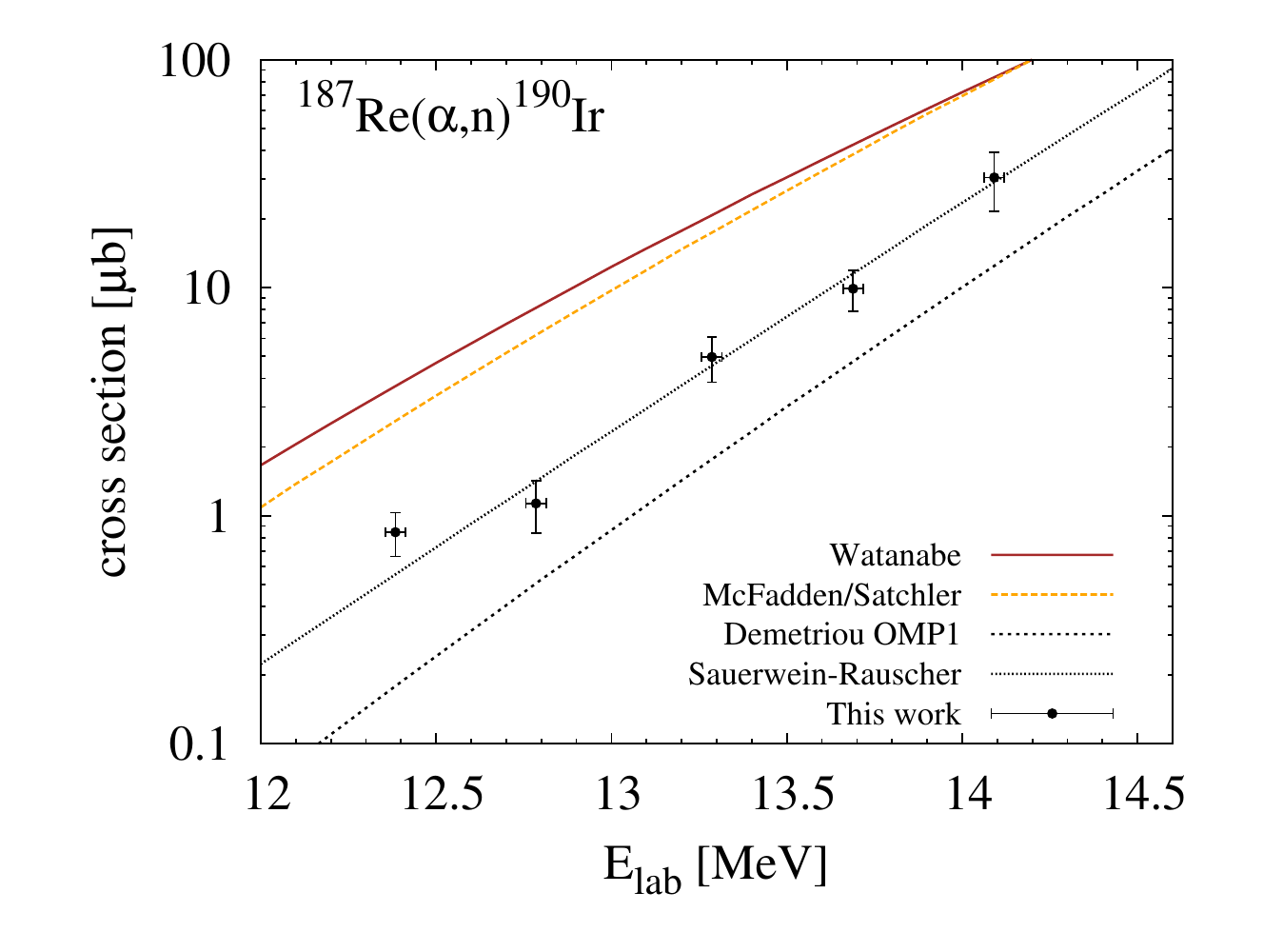}
\caption{(Color online) Total cross section as a function of the $\alpha$-particle energy in comparison to Statistical Model calculations using the TALYS code \cite{Talys}. The standard $\alpha$-OMP of TALYS is the one of Watanabe \cite{Watanabe} which overestimates the measured values by a factor of 4. Also the widely used $\alpha$-OMP of McFadden and Satchler \cite{McFadden} leads to a prediction of cross-section values which are too large. However, as pointed out in Ref. \cite{RauscherCoulex}, this may be due to the neglect of the Coulomb excitation in the determination of the $\alpha$-optical potential. In contrast to this, the optical-model potentials of Ref. \cite{Demetriou} lead to predictions of cross-section values which are too low (OMP1). The potential of Ref. \cite{Sauerwein} with a varied value for the ``steepness" of the Fermi-type function can reproduce the experimental values properly. See text for details.}
\label{fig:results}
\end{figure}

The \isotope[187]{Re}($\alpha$,n) reaction cross sections for $\alpha$-particle energies between 12.4 MeV and 14.1 MeV range from 0.85 $\mu$b to 30.4 $\mu$b. The uncertainties given for the cross-section values for each $\gamma$-ray transition as well as for the weighted means include systematical uncertainties of the photo-peak efficiency as well as the uncertainties for the number of $\alpha$-particles, the number of target nuclei, and the half-life ($\approx$~10~\%).  The results are given in Table \ref{tab:results} and Fig. \ref{fig:results}. The rather large uncertainties for the final values can be traced back partly to the determination of the $\eta$ parameters and the low intensity in the $\gamma$-ray spectra and partly to the large summing correction. 

As mentioned earlier, the \isotope[187]{Re}($\alpha$,n) cross section is almost only sensitive to the $\alpha$ width in the measured energy range. Thus, deviations of theoretical predictions obtained by Statistical Model calculations from the measured values are mainly caused by the accuracy of the adopted model for the $\alpha$-OMP. 

Therefore, calculations using the Statistical Model code TALYS (v1.4) \cite{Talys} were performed for different $\alpha$-OMPs. By default, TALYS provides different implementations for $\alpha$-OMPs.  The standard $\alpha$-OMP of TALYS is a potential derived from Watanabe \cite{Watanabe}. The results of Statistical Model calculations using this  $\alpha$-OMP fail to describe the measured data properly, as can be seen in Fig. \ref{fig:results}. The drastic overestimation of the cross-section values results in an $\chi^2$ of 285. 

Other available  $\alpha$-OMPs of TALYS are different versions of the  $\alpha$-OMP described in Ref. \cite{Demetriou} which is a global semi-microscopic approach for an $\alpha$-OMP (in Fig. \ref{fig:results} named OMP1). The results of TALYS using these  $\alpha$-OMPs are much better compared to the ones using the  $\alpha$-OMPs of Watanabe with $\chi^2$ between 2.7 and 5.6. Nevertheless, the predicted values are too low to reproduce the experimental values.

The  $\alpha$-OMP of McFadden and Satchler \cite{McFadden} was derived from extensive elastic $\alpha$-scattering data at comparably high energies. Therefore, it predicts elastic scattering and reaction data at high energies quite well but often seems to fail to reproduce experimental values at low energies \cite{Sauerwein} mainly due to the energy-independent imaginary part of the potential (see below). This is also the case for the \isotope[187]{Re}($\alpha$,n) reaction as one can see in Fig. \ref{fig:results} ($\chi^2$=143).
The cross-section values derived from Statistical Model calculations using the  $\alpha$-OMP by McFadden and Satchler overpredict the experimental values by a factor of 4. It was recently proposed by T. Rauscher, that  deviations of the $\alpha$-OMP of McFadden and Satchler at sub-Coulomb energies could be explained by the neglect of another inelastic channel \cite{RauscherCoulex}. A renormalization of the compound-formation cross section using the Coulomb excitation cross section led to a significant modification towards a smaller cross section for \isotope[144]{Sm}($\alpha$,$\gamma$)\isotope[148]{Gd} \cite{Somorjai}, \isotope[141]{Pr}($\alpha$,n)\isotope[144]{Pm} \cite{Sauerwein}, and \isotope[169]{Tm}($\alpha$,n)\isotope[172]{Lu} \cite{RauscherTm}.  Following the argumentation of Ref. \cite{RauscherCoulex}, the Coulomb excitation cross section should not be negligible in comparison to the compound-formation cross section in this case and should be subject to further investigations to confirm this claim. 


Somorjai \textit{et al.} have already pointed out another approach to modify the $\alpha$-OMP to improve the reproduction of the experimentally obtained \isotope[144]{Sm}($\alpha$,$\gamma$)\isotope[148]{Gd} reaction cross section \cite{Somorjai}. They argued that the depth of the imaginary part of the potential has to decrease with decreasing energy. This was achieved by parametrizing the depth of the imaginary part by an energy-dependent Fermi-type function. Sauerwein \textit{et al.} picked up the idea to improve the reproduction of \isotope[141]{Pr}($\alpha$,n)\isotope[144]{Pm} cross-section values by Statistical Model calculations \cite{Sauerwein}. One important requirement for their $\alpha$-OMP was to reduce the number of parameters. The so-called Sauerwein-Rauscher potential was thus based on the parametrization of Ref. \cite{McFadden} and only the depth of the imaginary part was modified by a Fermi-type function:
\begin{align}
W=\frac{W_0}{1+exp\left(\frac{0.9E_C-E_{c.m.}}{a}\right)}
\end{align}
Hence, the depth of the imaginary part depends on the height of the Coulomb barrier $E_C$, the center-of-mass energy $E_{c.m.}$ and only one parameter $a$ for the steepness of the energy dependence. In Ref. \cite{Sauerwein} as well as in Ref. \cite{Somorjai}, the steepness parameter was chosen to be $a=2$ MeV in order to give the best reproduction. Later, as the Sauerwein-Rauscher potential was tested on experimental data of $\alpha$-capture reaction cross-section values on \isotope[169]{Tm}, the best reproduction was given for steepness values of $a=4-6$ MeV \cite{RauscherTm}. Although the $\alpha$-OMP of Ref. \cite{McFadden} gave the best description of the measured cross section of the $\alpha$-capture reactions on \isotope[168]{Yb}, the results of the Sauerwein-Rauscher potential for $a = 4-6$ MeV were very promising \cite{Netterdon}. 
Motivated by its success, Statistical Model calculations using the Sauerwein-Rauscher potential were also performed for the \isotope[187]{Re}($\alpha$,n) reaction. Varying the steepness values between $a=4-6$ MeV resulted in an excellent reproduction of the experimental values. In Fig. \ref{fig:results}, the theoretical predictions based on the Sauerwein-Rauscher potential are shown for a steepness value of $a=4$ MeV ($\chi^2$=0.3). All the other parameters were kept the same as in Ref. \cite{Sauerwein}.

\section{Summary and Conclusion}
In this work, cross-section values for the \isotope[187]{Re}($\alpha$,n) reaction were measured at five different energies close to the astrophysically relevant energy region. The cross-section values for the lowest $\alpha$-particle energies have been extracted by means of $\gamma\gamma$-coincidence data.
Statistical Model calculations performed using the TALYS code (v1.4) revealed that the theoretical prediction of cross sections at sub-Coulomb energies is still problematic. Calculations based on the default implemented $\alpha$-OMPs of Ref. \cite{Watanabe, McFadden, Demetriou} show differences of the order of one magnitude. Due to the exclusive sensitivity of the reaction cross section on the $\alpha$-transmission, these variations can be traced back to the input for the $\alpha$-OMP. 
The success of the Sauerwein-Rauscher potential for this measurement as well as for previous ones \cite{Sauerwein, RauscherTm, Netterdon} is promising and should motivate further systematic studies of the global character of this $\alpha$-OMP. 

\begin{acknowledgments}
The authors acknowledge the help of A. Heiske and operators of PTB ion accelerator facility.
This project has been supported by the Deutsche Forschungsgemeinschaft under the contracts ZI 510/5-1 and INST 216/544-1 and the emerging group „ULDETIS“ within the UoC Excellence Initiative institutional strategy. 
\end{acknowledgments}


\begin{thebibliography}{38}%
\makeatletter
\providecommand \@ifxundefined [1]{%
 \@ifx{#1\undefined}
}%
\providecommand \@ifnum [1]{%
 \ifnum #1\expandafter \@firstoftwo
 \else \expandafter \@secondoftwo
 \fi
}%
\providecommand \@ifx [1]{%
 \ifx #1\expandafter \@firstoftwo
 \else \expandafter \@secondoftwo
 \fi
}%
\providecommand \natexlab [1]{#1}%
\providecommand \enquote  [1]{``#1''}%
\providecommand \bibnamefont  [1]{#1}%
\providecommand \bibfnamefont [1]{#1}%
\providecommand \citenamefont [1]{#1}%
\providecommand \href@noop [0]{\@secondoftwo}%
\providecommand \href [0]{\begingroup \@sanitize@url \@href}%
\providecommand \@href[1]{\@@startlink{#1}\@@href}%
\providecommand \@@href[1]{\endgroup#1\@@endlink}%
\providecommand \@sanitize@url [0]{\catcode `\\12\catcode `\$12\catcode
  `\&12\catcode `\#12\catcode `\^12\catcode `\_12\catcode `\%12\relax}%
\providecommand \@@startlink[1]{}%
\providecommand \@@endlink[0]{}%
\providecommand \url  [0]{\begingroup\@sanitize@url \@url }%
\providecommand \@url [1]{\endgroup\@href {#1}{\urlprefix }}%
\providecommand \urlprefix  [0]{URL }%
\providecommand \Eprint [0]{\href }%
\@ifxundefined \urlstyle {%
  \providecommand \doi  [0]{\begingroup \@sanitize@url \@doi}%
  \providecommand \@doi [1]{\endgroup \@@startlink {\doibase
  #1}doi:\discretionary {}{}{}#1\@@endlink }%
}{%
  \providecommand \doi  [0]{doi:\discretionary{}{}{}\begingroup
  \urlstyle{rm}\Url }%
}%
\providecommand \doibase [0]{http://dx.doi.org/}%
\providecommand \Doi [0]{\begingroup \@sanitize@url \@Doi }%
\providecommand \@Doi  [1]{\endgroup\@@startlink{\doibase#1}\@@Doi}%
\providecommand \@@Doi [1]{#1\@@endlink}%
\providecommand \selectlanguage [0]{\@gobble}%
\providecommand \bibinfo  [0]{\@secondoftwo}%
\providecommand \bibfield  [0]{\@secondoftwo}%
\providecommand \translation [1]{[#1]}%
\providecommand \BibitemOpen [0]{}%
\providecommand \bibitemStop [0]{}%
\providecommand \bibitemNoStop [0]{.\EOS\space}%
\providecommand \EOS [0]{\spacefactor3000\relax}%
\providecommand \BibitemShut  [1]{\csname bibitem#1\endcsname}%

\bibitem{woosley} S. E. Woosley, W. M. Howard, Astrophys. J. Suppl. \textbf{36} (1978) 285.
\bibitem{rayet} M. Rayet, M. Arnould, M. Hashimoto, N. Prantzos, K. Nomoto, Astron. Astroph. \textbf{298} (1995) 517.
\bibitem{nemeth} Zs. N\^{e}meth, F. K\"appeler, C. Theis, T. Belgya, S. W. Yates, Astrophys. J. \textbf{426} (1994) 357.
\bibitem{arlandini} C. Arlandini, F. K\"appeler, K. Wisshak, R. Gallino, M. Lugaro, M. Busso, O. Straniero, Astrophys. J. \textbf{525} (1999) 886.
\bibitem{arnould} M. Arnould, S. Goriely, Phys. Rep. \textbf{384} (2003) 1.
\bibitem{rauscherRep} T. Rauscher, N. Dauphas, I. Dillmann, C. Fr\"ohlich, Zs. F\"ul\"op, and Gy. Gy\"urky, Rep. Prog. Phys. \textbf{76} (2013) 066201.
\bibitem{Rauscher2002} T. Rauscher, A. Heger, R. D. Hoffman, S. E. Woosley, Astrophys. J. \textbf{576} (2002) 323.
\bibitem{Woosley2007} S. E. Woosley, A. Heger, Phys. Rep. \textbf{442} (2007) 269.
\bibitem{Travaglio2011} C. Travaglio, F. K. R\"opke, R. Gallino, W. Hillebrandt, Astrophys. J. \textbf{739} (2011) 93.
\bibitem{Hauser1952} W. Hauser, H. Feshbach, Phys. Rev. \textbf{87} (1952) 366.
\bibitem{Rauscher2000} T. Rauscher, F.-K. Thielemann, At. Data Nucl. Data Tables \textbf{75} (2000) 1.
\bibitem{Rauscher2011} T. Rauscher, Int. J. Mod. Phys. E \textbf{20} (2011) 5.

\bibitem{Kiss2011a} G. G. Kiss, T. Rauscher T. Sz\"ucs, Zs. Kert\^{e}sz, Gy. Gy\"urky, C. Fr\"ohlich, J. Farkas, Z. Elekes, E. Somorjai, Phys. Lett. B \textbf{695} (2011) 419.
\bibitem{Yalcin2009} C. Yal\c{c}in, R. T. G\"uray, N. \"Ozkan, S. Kutlu, Gy. Gy\"urky, J. Farkas, G. G. Kiss, Zs. F\"ul\"op, A. Simon, E. Somorjai, T. Rauscher, Phys. Rev. C \textbf{79} (2009) 065801.
\bibitem{Harissopulos2005} S. Harissopulos, A. Spyrou, A. Lagoyannis, Ch. Zarkadas, H.-W. Becker, C. Rolfs, F. Strieder, J. W. Hammer, A. Dewald, K.-O. Zell, P. von Brentano, R. Julin, P. Detriou, S. Goriely, Nucl. Phys. A \textbf{758} (2005) 505.
\bibitem{Gyurky2006} Gy. Gy\"urky, G. G. Kiss, Z. Elekes, Zs. F\"ul\"op, E. Somorjai, A. Palumbo, J. G\"orres, H. Y. Lee, W. Rapp, M. Wiescher, N. \"Ozkan, R. T. G\"uray, G. Efe, T. Rauscher, Phys. Rev. C \textbf{74} (2006) 025805.
\bibitem{Gyurky2010} Gy. Gy\"urky, Z. Elekes, J. Farkas, Zs. F\"ul\"op, Z. Hal\^{a}sz, G. G. Kiss, E. Somorjai, T. Sz\"ucs, R. T. G\"uray, N. \"Ozkan, C. Yal\c{c}in, T. Rauscher, J. Phys. G \textbf{37} (2010) 115201.
\bibitem{Rapp2002} W. Rapp, M. Heil, D. Hentschel, F. K\"appeler, R. Reifarth, H. J. Brede, H. Klein, T. Rauscher, Phys. Rev. C \textbf{66} (2002) 015803.
\bibitem{Rapp2008} W. Rapp, I. Dillmann, F. K\"appeler, U. Giesen, H. Klein, T. Rauscher, D. Hentschel, S. Hilpp, Phys. Rev. C \textbf{78} (2008) 025804.
\bibitem{Sauerwein} A. Sauerwein, H.-W. Becker, H. Dombrowski, M. Elvers, J. Endres, U. Giesen, J. Hasper, A. Hennig, L. Netterdon, T. Rauscher, D. Rogalla, K. O. Zell, and A. Zilges, Phys. Rev. C \textbf{84}, 045808 (2011).
\bibitem{Avrigeanu2010} M. Avrigeanu and V. Avrigeanu, Phys. Rev. C \textbf{82} (2010) 014606.
\bibitem{sensi} T. Rauscher, Astrophys. J. Suppl. \textbf{201} 26 (2012).
\bibitem{NNDC} NNDC Online Data Service, ENSDF database, http://www.nndc.bnl.gov/nndc/ensdf/.
\bibitem{brede} H. J. Brede, M. Cosack, G. Dietze, H. Gumpert, S. Guldbakke, R. Jahr, M. Kutscha, D. Schlegel-Bickmann, H. Sch\"olermann, Nucl. Instr. Meth. A \textbf{169}, 349 (1999).
\bibitem{boettge}  R. B\"ottger (private communication).
\bibitem{ziegler} J. Ziegler, J. Biersack, M. Ziegler, \textit{SRIM - The Stopping and Range of Ions in Matter}.
\bibitem{lise++} O. Tarasov, D. Bazin, M. Lewitowicz, O. Sorlin, Nucl.\ Phys.\ A\ \textbf{701}, $p.$ 661-665 (2002).
\bibitem{Netterdon} L. Netterdon, P. Demetriou, J. Endres, U. Giesen, G. G. Kiss, A. Sauerwein, T. Sz\"ucs, K. O. Zell, and A. Zilges, Nucl. Phys. A \textbf{916}, 149-167 (2013).


\bibitem{xia1} B. Hubbard-Nelson, M. Momayezi, and W. Warburton, Nucl. Instr. Meth. A \textbf{422}, 411 (1999).
\bibitem{xia2} W. Skulski, M. Momayezi, B. Hubbard-Nelson, P. Grudberg, J. Harris, and W. Warburton, Acta Phys. Pol. B \textbf{31}, 47 (2000).
\bibitem{geant4-1} S. Agostinelli, J. Allison, K. Amako, J. Apostolakis, H. Araujo, P. Arce, M. Asai, D. Axen, S. Banerjee, G. Barr $et\ al.$, Nucl. Instr. Meth. A, \textbf{503} 3, 250 (2003).
\bibitem{geant4-2} J. Allison, K. Amako, J. Apostolakis, H. Araujo, P. A. Dubois, M. Asai, G. Barr, R. Capra, S. Chauvie, R. Chytracek $et\ al.$, Nuclear Science and IEEE Transactions on Nuclear Science, \textbf{53} 1, 270 (2006).

\bibitem{Duchene} G. Duch\^{e}ne, F. A. Beck, P. J. Twin, G. de France, D. Curien, L. Han, C. W. Beausan, M. A. Bentley, P. J. Nolan, J. Simpson, Nucl. Instr. Meth. A, \textbf{432}, 90 (1999).
\bibitem{Talys} A. Koning, S. Hilaire, M. Duijvestijn, presented at the Proceedings of the International Conference
on Nuclear Data for Science and Technology, April 22-27, 2007, Nice, France, ed. by O. Bersillon, F. Gunsing, E. Bauge, R. Jacqmin, S. Leray, pp. 211–214.
\bibitem{Watanabe} S. Watanabe, Nucl. Phys. \textbf{8}, 484-492 (1958).
\bibitem{McFadden} L. McFadden, G. R. Satchler, Nucl. Phys. \textbf{84}, 177-200 (1966).
\bibitem{Demetriou} P. Demetriou, C. Grama, S. Goriely, Nucl. Phys. A \textbf{707}, 253–276 (2002).
\bibitem{RauscherCoulex} T. Rauscher, Phys. Rev. Lett. \textbf{111}, 061104 (2013).
\bibitem{Somorjai} E. Somorjai, Zs. F\"ul\"op, \'{A}. Z. Kiss, C. E. Rolfs, H. P. Trautvetter, U. Greife, M. Junker, S. Goriely, M. Arnould, M. Rayet, T. Rauscher, and H. Oberhummer, Astron. Astrophys. \textbf{333}, 1112-1116 (1998).
\bibitem{RauscherTm} T. Rauscher, G. G. Kiss, T. Sz\"ucs, Zs. F\"ul\"op, C. Fr\"ohlich, Gy. Gy\"urky, Z. Hal\^{a}sz, Zs. Kert\^{e}sz, and E. Somorjai, Phys. Rev. C \textbf{86}, 015804 (2012).
\end{thebibliography}
%

\end{document}